# Statistical mechanics of neocortical interactions:
# High resolution path-integral calculation of short-term memory

Lester Ingber

*Lester Ingber Research, P.O. Box 857, McLean, Virginia 22101*

ingber@alumni.caltech.edu

and

Paul L. Nunez

*Department of Biomedical Engineering, Tulane University, New Orleans, Louisiana 70118*

pln@bmen.tulane.edu

We present high-resolution path-integral calculations of a previously developed model of short-term memory in neocortex. These calculations, made possible with supercomputer resources, supplant similar calculations made in L. Ingber, Phys. Rev. E **49**, 4652 (1994), and support coarser estimates made in L. Ingber, Phys. Rev. A **29**, 3346 (1984). We also present a current experimental context for the relevance of these calculations using the approach of statistical mechanics of neocortical interactions, especially in the context of electroencephalographic data.

PACS Nos.: 87.10.+e, 05.40.+j, 02.50.-r, 02.70.-c

## I. INTRODUCTION

This paper describes a higher-resolution calculation of a similar calculation performed in a recent paper [1], using supercomputer resources not available at that time, and are of the quality of resolution presented in a different system using the same path-integral code PATHINT [2]. A more detailed description of the theoretical basis for these calculations can be found in that paper, and in previous papers in this series of statistical mechanics of neocortical interactions (SMNI) [3-18].

The SMNI approach is to develop mesoscopic scales of neuronal interactions at columnar levels of hundreds of neurons from the statistical mechanics of relatively microscopic interactions at neuronal and synaptic scales, poised to study relatively macroscopic dynamics at regional scales as measured by scalp electroencephalography (EEG). Relevant experimental data are discussed in the SMNI papers at the mesoscopic scales, e.g., as in this paper's calculations, as well as at macroscopic scales of scalp EEG. Here, we demonstrate that the derived firings of columnar activity, considered as order parameters of the mesoscopic system, develop multiple attractors, which illuminate attractors that may be present in the macroscopic regional dynamics of neocortex.

The SMNI approach may be complementary to other methods of studying nonlinear neocortical dynamics at macroscopic scales. For example, EEG and magnetoencephalography data have been expanded in a series of spatial principal components (Karhunen-Loeve expansion). The coefficients in such expansions are identified as order parameters that characterize phase changes in cognitive studies [19,20] and epileptic seizures [21,22], which are not considered here.

The calculations given here are of minicolumnar interactions among hundreds of neurons, within a macrocolumnar extent of hundreds of thousands of neurons. Such interactions take place on time scales of several $\tau$, where $\tau$ is on the order of 10 msec (of the order of time constants of cortical pyramidal cells). This also is the observed time scale of the dynamics of short-term memory. We hypothesize that columnar interactions within and/or between regions containing many millions of neurons are responsible for phenomena at time scales of several seconds.

That is, the nonlinear evolution as calculated here at finer temporal scales gives a base of support for the phenomena observed at the coarser temporal scales, e.g., by establishing mesoscopic attractors at many macrocolumnar spatial locations to process patterns at larger regions domains. This motivates us to continue using the SMNI approach to study minicolumnar interactions across macrocolumns and across regions. For example, this could be approached with a mesoscopic neural network using a confluence of techniques drawn from SMNI, modern methods of functional stochastic calculus defining nonlinear Lagrangians, adaptive simulated annealing (ASA) [23], and parallel-processing computation, as previously reported [16]. Other developments of SMNI, utilizing coarser statistical scaling than presented here, have been used to more directly interface with EEG phenomena, including the spatial and temporal filtering observed experimentally [14,15,17,18].

Section II presents a current experimental and theoretical context for the relevance of these calculations. We stress that neocortical interactions take place at multiple local and global scales and that a confluence of experimental and theoretical approaches across these scales very likely will be required to improve our understanding of the physics of neocortex.

Section III presents our current calculations, summarizing 10 CPU days of Convex 120 supercomputer resources in several figures. These results support the original coarser arguments given in SMNI papers a decade ago [6,8].

Section IV presents our conclusions.

## II. EXPERIMENTAL AND THEORETICAL CONTEXT

### A. EEG studies

EEG provides a means to study neocortical dynamic function at the millisecond time scales at which information is processed. EEG provides information for cognitive scientists and medical doctors. A major challenge for this field is the integration of these kinds of data with theoretical and experimental studies of the dynamic structures of EEG.

Theoretical studies of the neocortical medium have involved local circuits with postsynaptic potential delays [24-27], global studies in which finite velocity of action potential and periodic boundary conditions are important [28-31], and nonlinear nonequilibrium statistical mechanics of neocortex to deal with multiple scales of interaction [3-18]. The local and the global theories combine naturally to form a single theory in which control parameters effect changes between more local and more global dynamic behavior [31,32], in a manner somewhat analogous to localized and extended wave-function states in disordered solids.

Recently, plausible connections between the multiple-scale statistical theory and the more phenomenological global theory were proposed [14]. Experimental studies of neocortical dynamics with EEG include maps of magnitude distribution over the scalp [29,33], standard Fourier analyses of EEG time series [29], and estimates of correlation dimension [34,35]. Other studies have emphasized that many EEG states are accurately described by a few coherent spatial modes exhibiting complex temporal behavior [19-22,29,31]. These modes are the order parameters at macroscopic scales that underpin the phase changes associated with changes of physiological state.

The recent development of methods to improve the spatial resolution of EEG has made it more practical to study spatial structure. The new high resolution methods provide apparent resolution in the 2-3 cm range, as compared to 5-10 cm for conventional EEG [36]. EEG data were obtained in collaboration with the Swinburne Centre for Applied Neurosciences using 64 electrodes over the upper scalp. These scalp data are used to estimate potentials at the neocortical surface. The algorithms make use of general properties of the head volume conductor. A straightforward approach is to calculate the surface Laplacian from spline fits to the scalp potential distribution. This approach yields estimates similar to those obtained using concentric spheres models of the head [36].

Here we report on data recorded from one of us (P.L.N), while awake and relaxed with closed eyes (the usual alpha rhythm). The resulting EEG signal has dominant power in the 9-10 Hz range. We Fourier transformed the 64 data channels and passed Fourier coefficients at 10 Hz through our Laplacian algorithm to obtain cortical Fourier coefficients. In this manner the magnitude and phase structure of EEG was estimated. A typical Laplacian magnitude and phase plot for 1 sec of EEG is shown in Fig. 1. This structure was determined to be stable on 1-min time scales; that is averages over 1 min exhibit minimal minute to minute changes when the psychological/physiological state of the brain is held fixed. By contrast, the structure is quasi-stable on 1-sec time scales. To show this we calculated magnitude and phase templates based on an average over 3 min. We than obtained correlation coefficients by comparing magnitudes and phases at each electrode position for one second epochs of data with the templates. In this manner we determined that the structure is quasi-stable on 1 sec time scales. That is, correlation coefficients vary from second to second over moderate ranges, as shown in Fig. 2. Another interesting aspect of these data is the periodic behavior of the correlation coefficients; magnitudes and phases undergo large changes roughly every 6 sec and then return to patterns that more nearly match templates.

We have previously considered how mesoscopic activity may influence the very large scale dynamics observed on the scalp [14]. In some limiting cases (especially those brain states with minimal cognitive processing), this mesoscopic influence may be sufficiently small so that macroscopic dynamics can be approximated by a quasi-linear "fluid-like" representation of neural mass action [28-31]. In this approximation, the dynamics is crudely described as standing waves in the closed neocortical medium with periodic boundary conditions. Each spatial mode may exhibit linear or limit cycle behavior at frequencies in the 2–20 Hz range with mode frequencies partly determined by the size of the cortex and the action potential velocity in corticocortical fibers. The phase structure shown in Fig. 1 may show the nodal lines of such standing waves.

### B. Short-term memory

SMNI has presented a model of short-term memory (STM), to the extent it offers stochastic bounds for this phenomena during focused selective attention [1,6,8,37-39], transpiring on the order of tenths of a second to seconds, limited to the retention of $7 \pm 2$ items [40]. This is true even for apparently exceptional memory performers who, while they may be capable of more efficient encoding and retrieval of STM, and while they may be more efficient in "chunking" larger patterns of information into single items, nevertheless are limited to a STM capacity of $7 \pm 2$ items [41]. Mechanisms for various STM phenomena have been proposed across many spatial scales [42]. This "rule" is verified for acoustical STM, but for visual or semantic STM, which typically require longer times for rehearsal in an hypothesized articulatory loop of individual items, STM capacity appears to be limited to $4 \pm 2$ [43].

Another interesting phenomenon of STM capacity explained by SMNI is the primacy versus recency effect in STM serial processing, wherein first-learned items are recalled most error-free, with last-learned items still more error-free than those in the middle [44]. The basic assumption being made is that a pattern of neuronal firing that persists for many $\tau$ cycles is a candidate to store the "memory" of activity that gave rise to this pattern. If several firing patterns can simultaneously exist, then there is the capability of storing several memories. The short-time probability distribution derived for the neocortex is the primary tool to seek such firing patterns.

It has been noted that experimental data on velocities of propagation of long-ranged fibers [29,31] and derived velocities of propagation of information across local minicolumnar interactions [4] yield comparable times scales of interactions across minicolumns of tenths of a second. Therefore, such phenomena as STM likely are inextricably dependent on interactions at local and global scales, and this is assumed here.

### III. PRESENT CALCULATIONS

#### A. Probability distribution and the Lagrangian

As described in more detail in a previous paper [1], the short-time conditional probability of changing firing states within relaxation time $\tau$ of excitatory ($E$) and inhibitory ($I$) firings in a minicolumn of 110 neurons (twice this number in the visual neocortex) is given by the following summary of equations. The Einstein summation convention is used for compactness, whereby any index appearing more than once among factors in any term is assumed to be summed over, unless otherwise indicated by vertical bars, e.g., $|G|$. The mesoscopic probability distribution $P$ is given by the product of microscopic probability distributions $p_{\sigma_i}$, constrained such that the aggregate mesoscopic excitatory firings $M^E = \sum_{j \in E} \sigma_j$, and the aggregate mesoscopic inhibitory firings $M^I = \sum_{j \in I} \sigma_j$.

$$P = \prod_{G=E,I} P^G[M^G(r; t+\tau) | M^{\bar{G}}(r'; t)]$$

$$= \sum_{\sigma_j} \delta\left(\sum_{j \in E} \sigma_j - M^E(r; t+\tau)\right) \delta\left(\sum_{j \in I} \sigma_j - M^I(r; t+\tau)\right) \prod_j^N p_{\sigma_j}$$

$$\approx \prod_G (2\pi\tau g^{GG})^{-1/2} \exp(-N\tau \underline{L}^G) , \qquad (1)$$

where the final form is derived using the fact that $N > 100$. $\bar{G}$ represents contributions from both $E$ and $I$ sources. This defines the Lagrangian, in terms of its first-moment drifts $g^G$, its second-moment diffusion matrix $g^{GG'}$, and its potential $\underline{V}'$, all of which depend sensitively on threshold factors $F^G$,

$$P \approx (2\pi\tau)^{-1/2} g^{1/2} \exp(-N\tau \underline{L}) ,$$

$$\underline{L} = (2N)^{-1}(\dot{M}^G - g^G)g_{GG'}(\dot{M}^{G'} - g^{G'}) + M^G J_G/(2N\tau) - \underline{V}' ,$$

$$\underline{V}' = \sum_G \underline{V}''^G_{G'}(\rho \nabla M^{G'})^2 ,$$

$$g^G = -\tau^{-1}(M^G + N^G \tanh F^G) ,$$

$$g^{GG'} = (g_{GG'})^{-1} = \delta^{G'}_G \tau^{-1} N^G \text{sech}^2 F^G ,$$

$$g = \det(g_{GG'}) ,$$

$$F^G = \frac{(V^G - a^{|G|}_{G'} v^{|G|}_{G'} N^{G'} - \frac{1}{2} A^{|G|}_{G'} v^{|G|}_{G'} M^{G'})}{\{\pi[(v^{|G|}_{G'})^2 + (\phi^{|G|}_{G'})^2](a^{|G|}_{G'} N^{G'} + \frac{1}{2} A^{|G|}_{G'} M^{G'})\}^{1/2}} ,$$

$$a^G_{G'} = \frac{1}{2} A^G_{G'} + B^G_{G'} , \tag{2}$$

where $A^G_{G'}$ and $B^G_{G'}$ are macrocolumnar-averaged interneuronal synaptic efficacies, $v^G_{G'}$ and $\phi^G_{G'}$ are averaged means and variances of contributions to neuronal electric polarizations, and nearest-neighbor interactions $\underline{V}'$ are detailed in other SMNI papers [4,6]. $M^{G'}$ and $N^{G'}$ in $F^G$ are afferent macrocolumnar firings, scaled to efferent minicolumnar firings by $N/N^* \sim 10^{-3}$, where $N^*$ is the number of neurons in a macrocolumn. Similarly, $A^{G'}_G$ and $B^{G'}_G$ have been scaled by $N^*/N \sim 10^3$ to keep $F^G$ invariant. This scaling is for convenience only. For neocortex, due to chemical independence of excitatory and inhibitory interactions, the diffusion matrix $g^{GG'}$ is diagonal.

The above development of a short-time conditional probability for changing firing states at the mesoscopic entity of a mesocolumn (essentially a macrocolumnar averaged minicolumn), can be folded in time over and over by path-integral techniques developed in the late 1970s to process multivariate Lagrangians nonlinear in their drifts and diffusions [45,46]. This is further developed in the SMNI papers into a full spatial-temporal field theory across regions of neocortex.

## B. PATHINT algorithm

The PATHINT algorithm can be summarized as a histogram procedure that can numerically approximate the path integral to a high degree of accuracy as a sum of rectangles at points $M_i$ of height $P_i$ and width $\Delta M_i$. For convenience, just consider a one-dimensional system. The path-integral representation described above can be written, for each of its intermediate integrals, as

$$P(M; t + \Delta t) = \int dM' [g_s^{1/2}(2\pi \Delta t)^{-1/2} \exp(-L_s \Delta t)] P(M'; t)$$

$$= \int dM' G(M, M'; \Delta t) P(M'; t) ,$$

$$P(M; t) = \sum_{i=1}^{N} \pi(M - M_i) P_i(t) ,$$

$$\pi(M - M_i) = \begin{cases} 1, & (M_i - \frac{1}{2}\Delta M_{i-1}) \leq M \leq (M_i + \frac{1}{2}\Delta M_i) , \\ 0, & \text{otherwise} . \end{cases} \quad (3)$$

This yields

$$P_i(t + \Delta t) = T_{ij}(\Delta t) P_j(t) ,$$

$$T_{ij}(\Delta t) = \frac{2}{\Delta M_{i-1} + \Delta M_i} \int_{M_i - \Delta M_{i-1}/2}^{M_i + \Delta M_i/2} dM \int_{M_j - \Delta M_{j-1}/2}^{M_j + \Delta M_j/2} dM' G(M, M'; \Delta t) . \quad (4)$$

$T_{ij}$ is a banded matrix representing the Gaussian nature of the short-time probability centered about the (possibly time-dependent) drift. Care must be used in developing the mesh in $\Delta M^G$, which is strongly dependent on the diagonal elements of the diffusion matrix, e.g.,

$$\Delta M^G \approx (\Delta t g^{|G||G|})^{1/2} . \quad (5)$$

Presently, this constrains the dependence of the covariance of each variable to be a nonlinear function of that variable, albeit arbitrarily nonlinear, in order to present a straightforward rectangular underlying mesh.

A previous paper [1] attempted to circumvent this restriction by taking advantage of previous observations [6,8] that the most likely states of the "centered" systems lie along diagonals in $M^G$ space, a line determined by the numerator of the threshold factor, essentially

$$A_E^E M^E - A_I^E M^I \approx 0 , \tag{6}$$

where for neocortex $A_E^E$ is on the order of $A_I^E$. Along this line, for a centered system, the threshold factor $F^E \approx 0$, and $\underline{L}^E$ is a minimum. However, looking at $\underline{L}^I$, in $F^I$ the numerator $(A_E^I M^E - A_I^I M^I)$ is typically small only for small $M^E$, since for the neocortex $A_I^I \ll A_E^I$.

### C. Further considerations for high-resolution calculation

However, several problems plagued these calculations. First, and likely most important, is that it was recognized that a Sun workstation was barely able to conduct tests at finer mesh resolutions. This became apparent in a subsequent calculation in a different system, which could be processed at finer and finer meshes, where the resolution of peaks was much more satisfactory [2]. Second, it was difficult, if not impossible given the nature of the algorithm discussed above, to disentangle any possible sources of error introduced by the approximations based on the transformation used.

The main issues to note here are that the physical boundaries of firings $M^G = \pm N^G$ are imposed by the numbers of excitatory and inhibitory neurons per minicolumn in a given region. Physically, firings at these boundaries are unlikely in normal brains, e.g., unless they are epileptic or dead. Numerically, PATHINT problems with SMNI diffusions and drifts arise for large $M^G$ at these boundaries:

(a) SMNI has regions of relatively small diffusions $g^{GG'}$ at the boundaries of $M^G$ space. As the $\Delta M^G$ meshes are proportional to $(g^{GG'}\Delta t)^{1/2}$, this could require PATHINT to process relatively small meshes in these otherwise physically uninteresting domains, leading to kernels of size tens of millions of elements. These small diffusions also lead to large Lagrangians which imply relatively small contributions to the conditional probabilities of firings in these domains.

(b) At the boundaries of $M^G$ space, SMNI can have large negative drifts, $g^G$. This can cause anomalous numerical problems with the Neumann reflecting boundary conditions taken at all boundaries. For example, if $g^G \Delta t$ is sufficiently large and negative, negative probabilities can result. Therefore, this would require quite small $\Delta t$ meshes to treat properly, affecting the $\Delta M^G$ meshes throughout $M^G$ space.

A quite reasonable solution is to cut off the drifts and diffusions at the edges by Gaussian factors $\Gamma$,

$$g^G \to g^G \Gamma ,$$

$$g^{GG'} \to g^{GG'}\Gamma + (1-\Gamma)N^G/\tau ,$$

$$\Gamma = \prod_{G=E,I} \frac{\exp[-(M^G/N^G)^2/C] - \exp(-1/C)}{1 - \exp(-1/C)} , \tag{7}$$

where $C$ is a cutoff parameter and the second term of the transformed diffusion is weighted by $N^G/\tau$, the value of the SMNI diffusion at $M^G = 0$. A value of $C = 0.2$ was found to give good results.

However, the use of this cutoff rendered the diffusions approximately constant over the $E$ and the $I$ firing states, e.g., on the order of $N^G$. Therefore, here the diffusions were taken to be these constants.

While a resolution of $\Delta t = 0.5\tau$ was taken for the previous PATHINT calculation [1], here a temporal resolution of $\Delta t = 0.01\tau$ was necessary to get well-developed peaks of the evolving distribution for time epochs on the order of several $\tau$. As discussed in the Appendix of an earlier paper [6], such a finer resolution is quite physically reasonable, i.e., even beyond any numerical requirements for such temporal meshes. That is, defining $\theta$ in that previous study to be $\Delta t$, firings of $M^G(t + \Delta t)$ for $0 \leq \Delta t \leq \tau$ arise due to interactions within memory $\tau$ as far back as $M^G(t + \Delta t - \tau)$. That is, the mesocolumnar unit expresses the firings of afferents $M^G(t + \tau)$ at time $t + \tau$ as having been calculated from interactions $M^G(t)$ at the $\tau$-averaged efferent firing time $t$. With equal likelihood throughout time $\tau$, any of the $N^*$ uncorrelated efferent neurons from a surrounding macrocolumn can contribute to change the minicolumnar mean firings and fluctuations of their $N$ uncorrelated minicolumnar afferents. Therefore, for $\Delta t \leq \tau$, at least to resolution $\Delta t \geq \tau/N$ and to order $\Delta t/\tau$, it is reasonable to assume that efferents effect a change in afferent mean firings of $\Delta t \dot{M}^G = M^G(t + \Delta t) - M^G(t) \approx \Delta t g^G$ with variance $\Delta t g^{GG}$. Indeed, columnar firings (e.g., as measured by averaged evoked potentials) are observed to be faithful continuous probabilistic measures of individual neuronal firings (e.g., as measured by poststimulus histograms) [47].

When this cutoff procedure is applied with this temporal mesh, an additional physically satisfying result is obtained, whereby the $\Delta M^G$ mesh is on the order of a firing unit throughout $M^G$ space. The interesting physics of the interior region as discussed in previous papers is still maintained by this procedure.

### D. Four models of selective attention

Three representative models of neocortex during states of selective attention are considered, which are effected by considering synaptic parameters within experimentally observed ranges.

A model of dominant inhibition describes how minicolumnar firings are suppressed by their neighboring minicolumns. For example, this could be effected by developing nearest-neighbor mesocolumnar interactions [5], but the averaged effect is established by inhibitory mesocolumns (IC) by setting $A_E^I = A_I^E = 2A_E^E = 0.01 N^*/N$. Since there appears to be relatively little $I$—$I$ connectivity, we set $A_I^I = 0.0001 N^*/N$. The background synaptic noise is taken to be $B_I^E = B_E^I = 2B_E^E = 10B_I^I = 0.002 N^*/N$. As minicolumns are observed to have ~110 neurons (the visual cortex appears to have approximately twice this density) [48] and as there appear to be a predominance of $E$ over $I$ neurons [29], we take $N^E = 80$ and $N^I = 30$. As supported by references to experiments in early SMNI papers, we take $N^*/N = 10^3$, $J_G = 0$ (absence of long-ranged interactions), $V^G = 10$ mV, $|v_{G'}^G| = 0.1$ mV, and $\phi_{G'}^G = 0.1$ mV. It is discovered that more minima of $\bar{L}$ are created, or "restored," if the numerator of $F^G$ contains terms only in $\bar{M}^G$, tending to center the Lagrangian about $\bar{M}^G = 0$. Of course, any mechanism producing more as well as deeper minima is statistically favored. However, this particular centering mechanism has plausible support: $M^G(t + \tau) = 0$ is the state of afferent firing with highest statistical weight. That is, there are more combinations of neuronal firings $\sigma_j = \pm 1$ yielding this state more than any other $M^G(t + \tau)$; e.g., $\sim 2^{N^G+1/2}(\pi N^G)^{-1/2}$ relative to the states $M^G = \pm N^G$. Similarly, $M^{*G}(t)$ is the state of efferent firing with highest statistical weight. Therefore, it is natural to explore mechanisms that favor common highly weighted efferent and afferent firings in ranges consistent with favorable firing threshold factors $F^G \approx 0$.

The centering effect of the IC model of dominant inhibition, labeled here as the IC′ model, is quite easy for the neocortex to accommodate. For example, this can be accomplished simply by readjusting the synaptic background noise from $B_E^G$ to $B_E'^G$,

$$B'^G_E = \frac{V^G - (\frac{1}{2} A^G_I + B^G_I) v^G_I N^I - \frac{1}{2} A^G_E v^G_E N^E}{v^G_E N^G} \qquad (8)$$

for both $G = E$ and $G = I$. This is modified straightforwardly when regional influences from long-ranged firings $M^{\ddagger E}$ are included [15]. In general, $B^G_E$ and $B^G_I$ (and possibly $A^G_E$ and $A^G_I$ due to actions of neuromodulators and $J_G$ or $M^{\ddagger E}$ constraints from long-ranged fibers) are available to force the constant in the numerator to zero, giving an extra degree(s) of freedom to this mechanism. (If $B'^G_E$ would be negative, this leads to unphysical results in the square-root denominator of $F^G$. Here, in all examples where this occurs, it is possible to instead find positive $B'^G_I$ to appropriately shift the numerator of $F^G$.) In this context, it is experimentally observed that the synaptic sensitivity of neurons engaged in selective attention is altered, presumably by the influence of chemical neuromodulators on postsynaptic neurons [49].

By this centering mechanism, the model $F^G_{\text{IC}'}$ is obtained

$$F^E_{\text{IC}'} = \frac{0.5\bar{M}^I - 0.25\bar{M}^E}{\pi^{1/2}(0.1\bar{M}^I + 0.05\bar{M}^E + 10.4)^{1/2}},$$

$$F^I_{\text{IC}'} = \frac{0.005\bar{M}^I - 0.5\bar{M}^E}{\pi^{1/2}(0.001\bar{M}^I + 0.1\bar{M}^E + 20.4)^{1/2}}. \qquad (9)$$

The other "extreme" of normal neocortical firings is a model of dominant excitation, effected by establishing excitatory mesocolumns (EC) by using the same parameters $\{ B^G_{G'}, v^G_{G'}, \phi^G_{G'}, A^I_I \}$ as in the IC model, but setting $A^E_E = 2A^I_E = 2A^E_I = 0.01 N^*/N$. Applying the centering mechanism to EC, $B'^E_I = 10.2$ and $B'^I_I = 8.62$. This yields

$$F^E_{\text{EC}'} = \frac{0.25\bar{M}^I - 0.5\bar{M}^E}{\pi^{1/2}(0.05\bar{M}^I + 0.10\bar{M}^E + 17.2)^{1/2}},$$

$$F^I_{\text{EC}'} = \frac{0.005\bar{M}^I - 0.25\bar{M}^E}{\pi^{1/2}(0.001\bar{M}^I + 0.05\bar{M}^E + 12.4)^{1/2}}. \qquad (10)$$

Now it is natural to examine a balanced case intermediate between IC and EC, labeled BC. This is accomplished by changing $A^E_E = A^I_E = A^E_I = 0.005 N^*/N$. Applying the centering mechanism to BC, $B'^E_E = 0.438$ and $B'^I_I = 8.62$. This yields

$$F^E_{BC'} = \frac{0.25\bar{M}^I - 0.25\bar{M}^E}{\pi^{1/2}(0.050\bar{M}^E + 0.050\bar{M}^I + 7.40)^{1/2}} ,$$

$$F^I_{BC'} = \frac{0.005\bar{M}^I - 0.25\bar{M}^E}{\pi^{1/2}(0.001\bar{M}^I + 0.050\bar{M}^E + 12.4)^{1/2}} . \quad (11)$$

A fourth model, similar to BC′, for the visual neocortex is considered as well, BC′_VIS, where $N^G$ is doubled.

$$F^E_{BC'\_VIS} = \frac{0.25\bar{M}^I - 0.25\bar{M}^E}{\pi^{1/2}(0.050\bar{M}^E + 0.050\bar{M}^I + 20.4)^{1/2}} ,$$

$$F^I_{BC'\_VIS} = \frac{0.005\bar{M}^I - 0.25\bar{M}^E}{\pi^{1/2}(0.001\bar{M}^I + 0.050\bar{M}^E + 26.8)^{1/2}} . \quad (12)$$

### E. Results of calculations

Models BC′, EC′, and IC′ were run at time resolutions of $\Delta t = 0.01\tau$, resulting in firing meshes of $\Delta M^E = 0.894427$ (truncated as necessary at one end point to fall within the required range of $\pm N^E$), and $\Delta M^I = 0.547723$. To be sure of accuracy in the calculations, off-diagonal spreads of firing meshes were taken as ±5. This lead to an initial four-dimensional matrix of $179 \times 110 \times 11 \times 11 = 2\,382\,490$ points, which was cut down to a kernel of $2\,289\,020$ points because the off-diagonal points did not cross the boundaries. Reflecting Neumann boundary conditions were imposed by the method of images, consisting of a point image plus a continuous set of images leading to an error function [50]. A Convex 120 supercomputer was used, but there were problems with its C compiler, so gcc version 2.60 was built and used. Runs across several machines, e.g., Suns, Dec workstations, and Crays, checked reproducibility of this compiler on this problem. It required about 17 CPU min to build the kernel, and about 0.45 CPU min for each $\Delta t$-folding of the distribution.

For model BC′_VIS, the same time resolution and off-diagonal range was taken, resulting in firing meshes of $\Delta^E = 1.26491$ and $\Delta M^I = 0.774597$, leading to a kernel of size 4 611 275 elements. It required about 34 CPU min to build the kernel, and about 0.90 CPU min for each $\Delta t$ folding of the distribution.

An initial $\delta$-function stimulus was presented at $M^E \approx M^I \approx 0$ for each model. The subsequent dispersion among the attractors of the systems gives information about the pattern capacity of this system. Data was printed every 100 foldings, representing the evolution of one unit of $\tau$. For run BC′, data were collected for up to $50\tau$, and for the other models data were collected up to $30\tau$.

As pointed out in Sec. II, long-ranged minicolumnar circuitry across regions and across macrocolumns within regions is quite important in the neocortex and this present calculation only represents a model of minicolumnar interactions within a macrocolumn. Therefore, only the first few $\tau$ foldings should be considered as having much physical significance.

Figure 3(a) shows the evolution of model BC′ after 100 foldings of $\Delta t = 0.01$, or one unit of relaxation time $\tau$. Note the existence of ten well developed peaks or possible trappings of firing patterns. The peaks more distant from the center of firing space would be even smaller if the actual nonlinear diffusions were used, since they are smaller at the boundaries, increasing the Lagrangian and diminishing the probability distribution. However, there still are two obvious scales. If both scales are able to be accessed then all peaks are available to process patterns, but if only the larger peaks are accessible, then the capacity of this memory system is accordingly decreased. This seems to be able to describe the "$7 \pm 2$" rule. Figure 3(b) shows the evolution after 500 foldings at $5\tau$; note that the integrity of the different patterns is still present. Figure 3(c) shows the evolution after 1000 foldings at $10\tau$; note the deterioration of the patterns. Figure 3(d) shows the evolution after 3000 foldings at $30\tau$; note that while the original central peak has survived, now most of the other peaks have been absorbed into the central peaks and the attractors at the boundary.

Figure 4(a) shows the evolution of model EC′ after 100 foldings of $\Delta t = 0.01$, or one unit of relaxation time $\tau$. Note that, while ten peaks were present at this time for model BC′, now there are only four well developed peaks, of which only two are quite strong. Figure 4(b) shows the evolution after 1000 foldings at $10\tau$; note that only the two previously prominent peaks are now barely distinguishable.

Figure 5(a) shows the evolution of model IC′ after 100 foldings of $\Delta t = 0.01$, or one unit of relaxation time $\tau$. While similar to model BC′, here too there are ten peaks within the interior of firing space. However, quite contrary to that model, here the central peaks are much smaller and therefore less likely than the middle and the outer peaks (the outer ones prone to being diminished if nonlinear diffusions were used, as commented on above), suggesting that the original stimulus pattern at the origin cannot be strongly contained. Figure 5(b) shows the evolution after 100 foldings at $10\tau$; note that only the attractors at the boundaries are still represented.

Figure 6(a) shows the evolution of model BC′_VIS after 100 foldings of $\Delta t = 0.01$, or one unit of relaxation time $\tau$. In comparison to model BC′, this model exhibits only six interior peaks, with three scales of relative importance. If all scales are able to be accessed, then all peaks are available to process patterns, but if only the larger peaks are accessible, then the capacity of this memory system is accordingly decreased. This seems to be able to describe the "4 ± 2" rule for visual memory. Figure 6(b) shows the evolution after 100 foldings at $10\tau$; note that these peaks are still strongly represented. Also note that now other peaks at lower scales are clearly present, numbering on the same order as in the BC′ model, as the strength in the original peaks dissipates throughout firing space, but these are much smaller and therefore much less probable to be accessed. As seen in Fig. 6c, similar to the BC′ model, by $15\tau$, only the original two large peaks remain prominent.

## IV. CONCLUSION

Experimental EEG results are available for regional interactions and the evidence supports attractors that can be considered to process short-term memory under conditions of selective attention. There are many models of nonlinear phenomena that can be brought to bear to study these results.

There is not much experimental data available for large-scale minicolumnar interactions. However, SMNI offers a theoretical approach, based on experimental data at finer synaptic and neuronal scales, that develops attractors that are consistent with short-term memory capacity. The duration and the stability of such attractors likely are quite dependent on minicolumnar circuitry at regional scales, and further study will require more intensive calculations than presented here [16].

We have presented a reasonable paradigm of multiple scales of interactions of the neocortex under conditions of selective attention. Presently, global scales are better represented experimentally, but the mesoscopic scales are represented in more detail theoretically. We have offered a theoretical approach to consistently address these multiple scales [14-16], and more a phenomenological macroscopic theory [28-32] that is more easily compared with macroscopic data. We expect that future experimental efforts will offer more knowledge of the neocortex at these multiple scales as well.

## ACKNOWLEDGMENTS

We acknowledge the use of the Tulane University Convex 120 supercomputer for all calculations presented in this paper. We thank Richard Silberstein of the Centre for Applied Neurosciences for recording PLN's EEG data.

**FIGURE CAPTIONS**

FIG. 1. Magnitude (upper) and phase (lower) at 9 Hz of 1 sec of alpha rhythm is shown. The plots represent estimates of potential on the cortical surface calculated from a 64 channel scalp recording (average center-to center electrode spacing of about 2.7 cm). The estimates of cortical potential wave forms were obtained by calculating spatial spline functions at each time slice to obtain analytic fits to scalp potential distributions. Surface Laplacian wave forms were obtained from second spatial derivatives (in the two surface tangent coordinates). Magnitude and phase were obtained from temporal Fourier transforms of the Laplacian wave forms. This particular Laplacian algorithm yields estimates of cortical potential that are similar to inverse solutions based on four concentric spheres modes of the head. The Laplacian appears to be robust with respect to noise and head model errors [36]. The dark and the lighter shaded regions are 90° out of phase, suggesting quasi-stable phase structure with regions separated by a few centimeters 180° out of phase (possible standing waves). Data recorded at the Swinburne Centre for Applied Neurosciences in Melbourne, Australia.

FIG. 2. Changes of alpha rhythm correlation coefficients based on comparisons of magnitude (solid line) and phase (dashed line) plots of successive 1-sec epochs of alpha rhythm compared with spatial templates based on averages over 3 min of data (similar to Fig. 1). The data show a quasistable structure with major changes in magnitude or phase about every 6 sec, after which the structure tends to return to the template structure.

FIG. 3. Model BC′: (a) the evolution at $\tau$, (b) the evolution at $5\tau$, (c) the evolution at $10\tau$, and (d) the evolution at $30\tau$.

FIG. 4. Model EC′: (a) the evolution at $\tau$ and (b) the evolution at $10\tau$.

FIG. 5. Model IC′: (a) the evolution at $\tau$ and (b) the evolution at $10\tau$.

FIG. 6. Model BC′_VIS: (a) the evolution at $\tau$, (b) the evolution at $10\tau$, and (c) the evolution at $15\tau$.

# Figure 1 upper

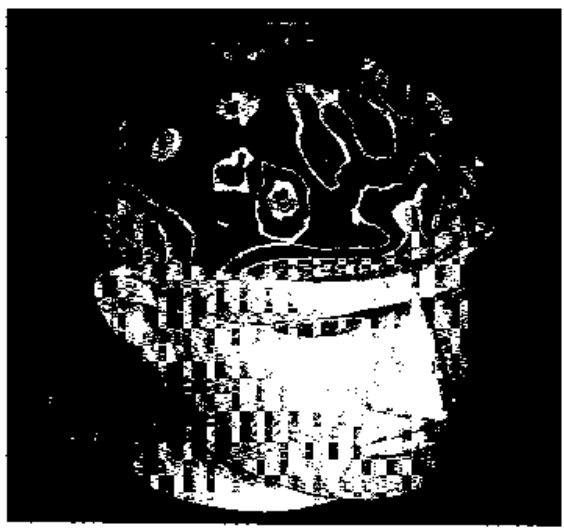

## Figure 1 lower

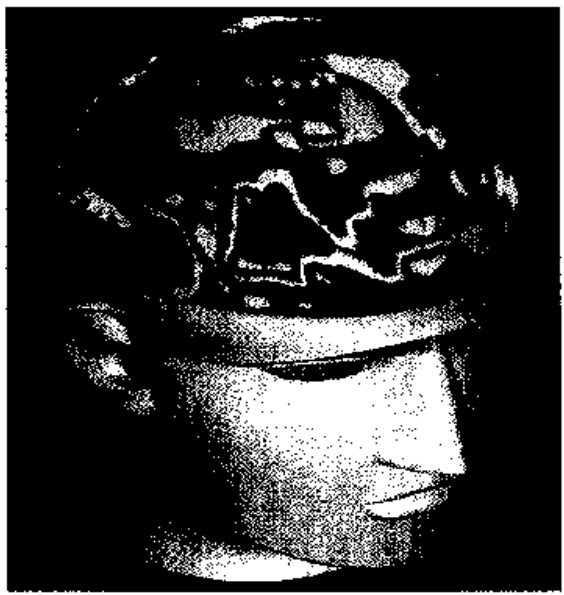

Figure 2

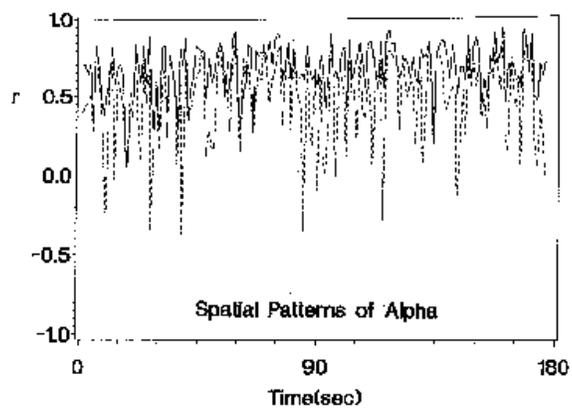

**Figure 3a**

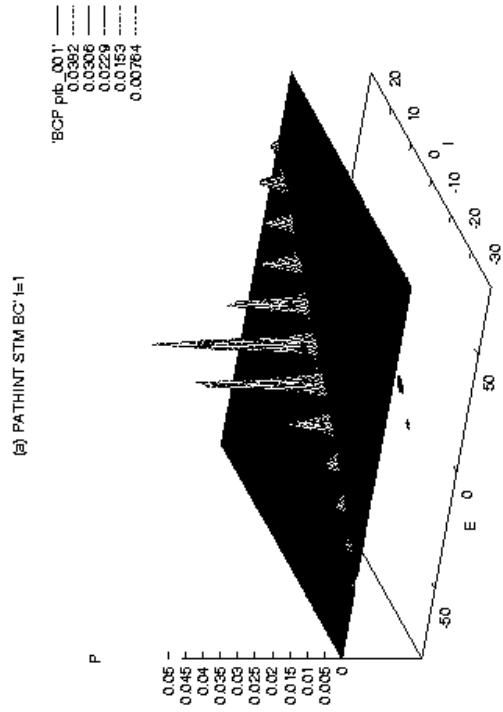

**Figure 3b**

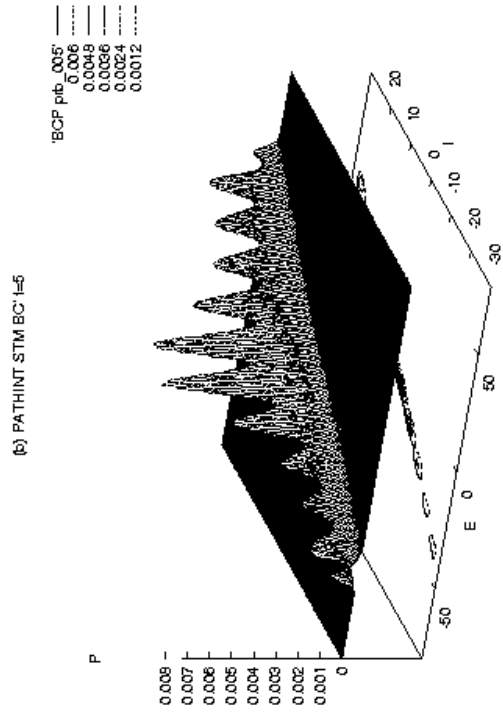

# Figure 3c

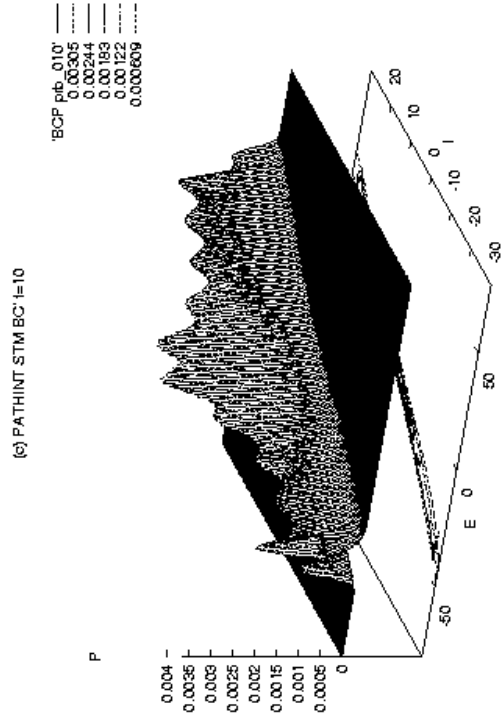

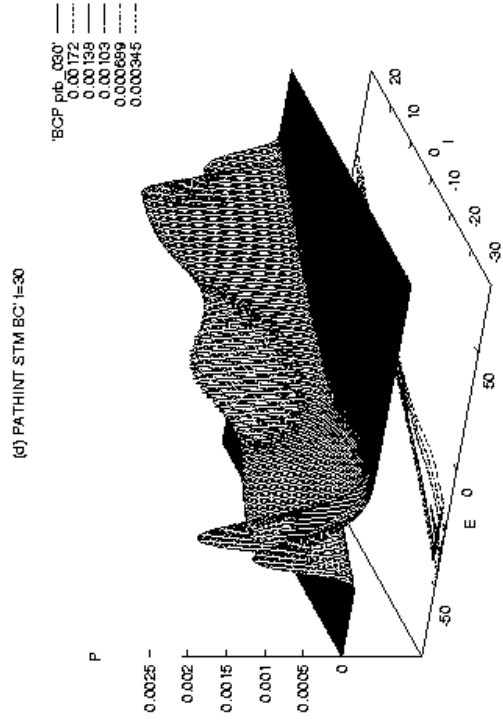

# Figure 4a

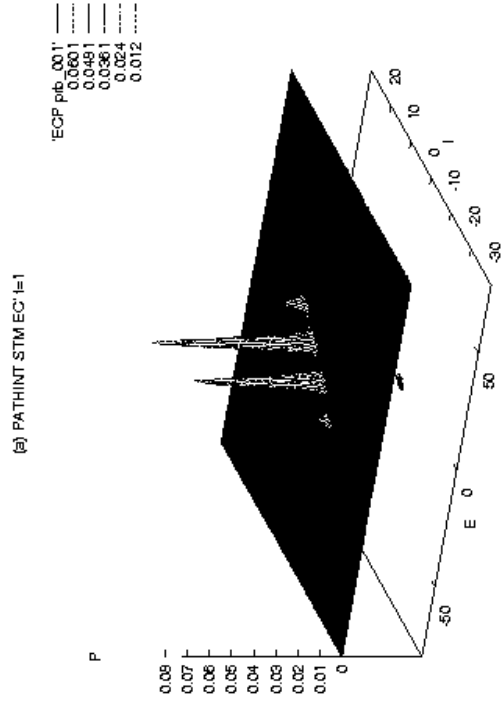

Figure 4b

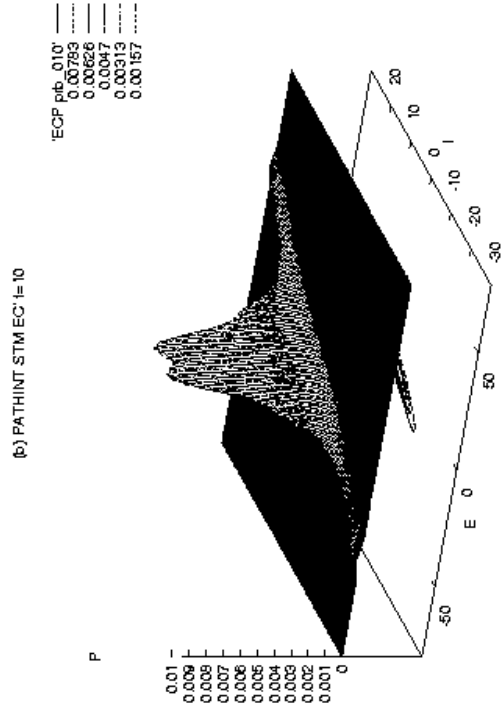

# Figure 5a

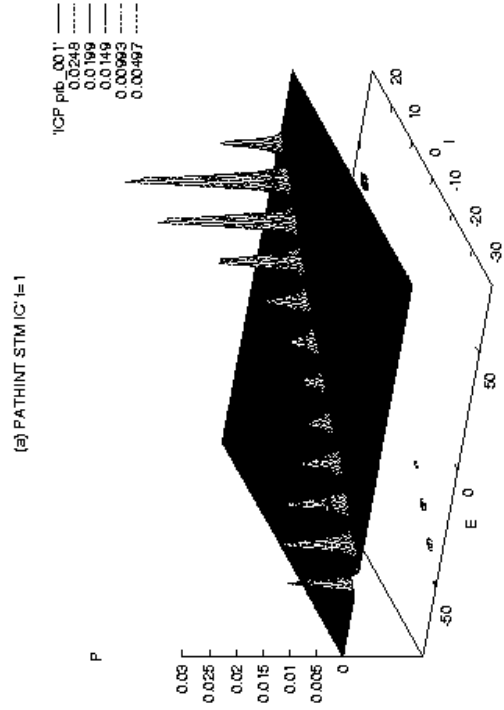

**Figure 5b**

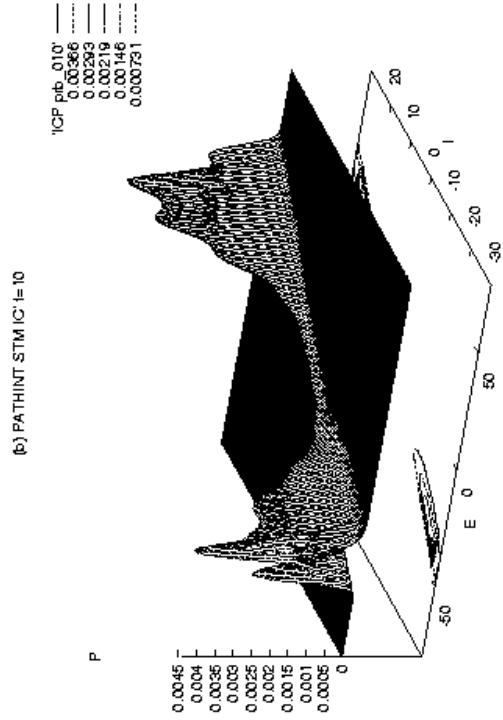

# Figure 6a

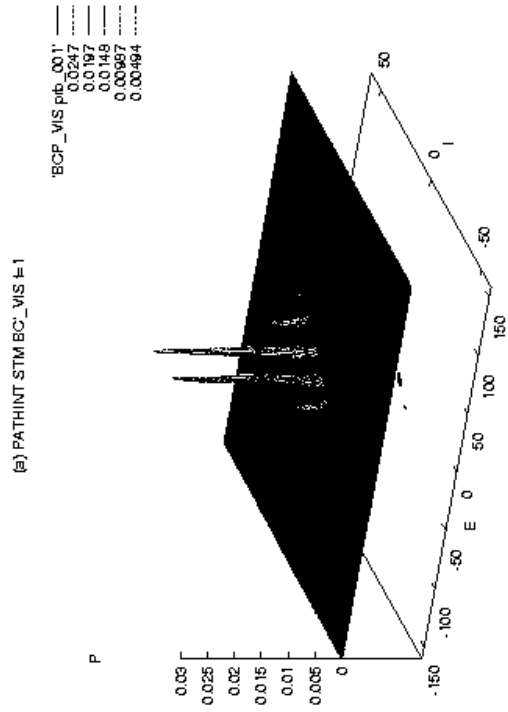

# Figure 6b

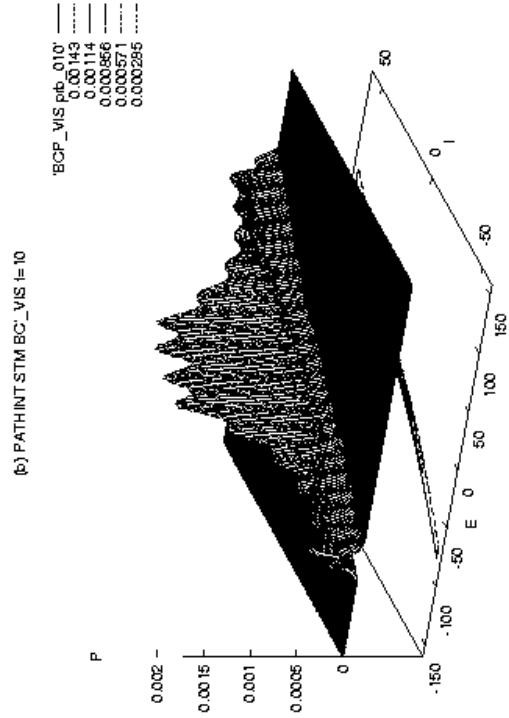

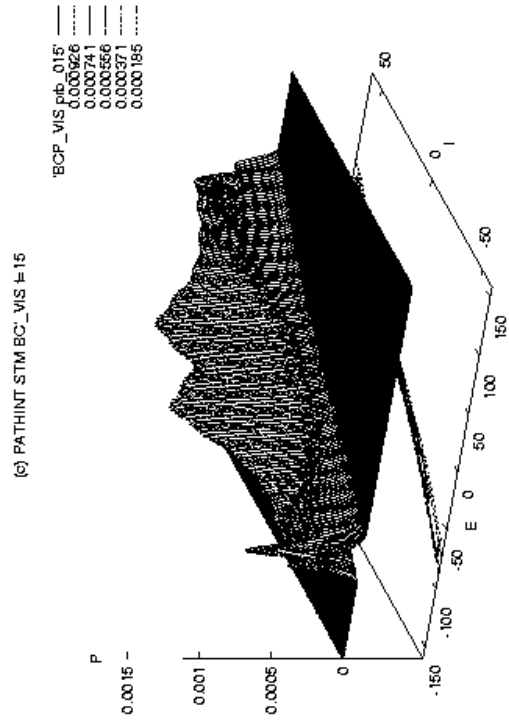